\documentclass[preprint,showpacs,preprintnumbers,amsmath,amssymb,nofootinbib]{revtex4}
\usepackage{graphicx}
\usepackage{bm}

\usepackage{epsfig}

\newcommand{\beq}{\begin{equation}}
\newcommand{\eeq}{\vspace{0cm} \end{equation}}
\newcommand{\beqq}{\setlength\arraycolsep{2pt}\begin{eqnarray}}
\newcommand{\eeqq}{\vspace{0cm} \end{eqnarray}}

\begin{document}

\title{On Phantom Thermodynamics}

\author{S. H. Pereira} \email{spereira@astro.iag.usp.br}
\author{J. A. S. Lima}\email{limajas@astro.iag.usp.br}

\affiliation{Departamento de Astronomia, Universidade de S\~{a}o
Paulo \\ Rua do Mat\~ao, 1226 - 05508-900, S\~ao Paulo, SP, Brazil}

\pacs{98.80.-k}
\keywords{dark energy, phantom thermodynamics}

\bigskip
\begin{abstract}
The thermodynamic properties of dark energy fluids described by an
equation of state parameter $\omega=p/\rho$ are rediscussed in the
context of  FRW type geometries.  Contrarily to previous claims, it
is argued here that  the phantom regime $\omega<-1$ is not physically 
possible since that both the temperature and the entropy of every 
physical fluids must be always positive definite.  This means
that one cannot appeal to negative temperature in order to save the
phantom dark energy hypothesis as has been recently done in the
literature.  Such a result remains true as long as the chemical
potential is zero. However, if the phantom fluid is endowed with a
non-null chemical potential, the phantom field hypothesis becomes
thermodynamically consistent, that is,  there are macroscopic
equilibrium states  with  $T>0$ and  $S>0$ in the course of the
Universe expansion. Further,  the negative value of the chemical 
potential resulting from the entropy constraint  ($S>0$) 
suggests a bosonic massless nature to the phantom particles.

\end{abstract}

\maketitle

\section{Introduction}

Several kinds of complementary astronomical observations indicate
that the Universe is expanding in an accelerated form and that the
transition (from a decelerating to an  accelerating regime) occurred
at a redshift of the order of unity \cite{SN,CMB}.  In the context
of general relativity, an accelerating stage and the associated
dimming of type Ia Supernovae  are usually explained by assuming the
existence of an exotic substance with negative pressure sometimes
called dark energy. Actually, for dark energy dominated models,  the
scale factor evolution  is governed by the equation 3${\ddot a}/a=
-4\pi G(\rho + 3p)$, and means that  a hypothetical component with
negative pressure satisfying $p< -\rho/3$ may  accelerate the
Universe ($\ddot a >0$).

There are many candidates to represent this extra non-luminous
relativistic component. In the case of XCDM cosmologies, for
instance, it can be phenomenologically  described by an  equation of
state (EoS) of the form \cite{review} 
\beq p=\omega
\rho,\label{eqstate} 
\eeq 
where $p$ and $\rho$ denotes  the pressure
and energy density, respectively, and  $\omega$ is a constant
negative parameter. The case $\omega = -1$ corresponds to a positive
cosmological constant, or vacuum energy, while for $\omega < -1$ we
have the so called phantom dark energy regime \cite{phantom}, or phantom 
fluids\footnote{Indeed, in the standard lines of the thermodynamic, 
we will see that it does not make sense to speak of phantom fluids for systems with null chemical potential.}.

The case for a phantom dominated Universe has been first suggested
with basis on SN-Ia analysis alone which favor $\omega < -1$ more than
cosmological constant or quintessence \cite{coras}. A more precise
observational data analysis (involving CMB, Hubble Space Telescope,
type Ia Supernovae, and 2dF data sets) allows the equation of state
$p=\omega\rho$ with a constant $\omega$ on the interval [-1.38,
-0.82] at the 95\% C. L. \cite{Melc}.

>From a theoretical point of view, the study of phantom regime is
also a very interesting subject mainly due to a long list of
pathologies. Initially, it was criticized by several authors due to
issues of stability \cite{pg12} which must be added to
some weird properties, like the possibility of superluminal sound
speed,  as well as the violation of some classical energy conditions
\cite{pg4}. In particular, since ($p+\rho < 0$) one may see that it
violates the strong and dominant energy conditions. Further, the
energy density of a phantom field increases along the cosmic
evolution thereby  causing a super accelerating universe which will
end in a doomsday state dubbed {\it Big Rip} \cite{Cald} which is of
type I singularity according to the Barrow classification scheme
\cite{Barrow}.  Such a {\it Big Rip} singularity corresponds to
$\rho \rightarrow \infty$ at a finite time in the future which presumably 
will be avoided only if 
one considers possible effects from quantum gravity. A quantum treatment on the phantom regime has been discussed by several authors \cite{onemli}.

Another interesting point concerns the study of the spectral
distribution and some related thermodynamical properties of the
phantom fluid, like their temperature and entropy. We have two different 
approaches to study the thermodynamic of phantom fluids. The first, based
on a somewhat ambiguous thermodynamic deduction \cite{donam} (see
discussion in section II), was given by Gonz\'alez-D\'{i}az and  C. L. Sig\"{u}enza \cite{gonzalez}, 
which claimed that the temperature  of phantom
fluids in a Friedmann-Robertson-Walker (FRW) geometry should be
negative and defined by the scaling law 
\beq 
T \sim (1+\omega)a^{-3\omega}\,,\label{temp1} 
\eeq 
where $a(t)$ is the scale
factor (note the negative prefactor, $1 + \omega$, multiplying the
power of the FRW scale function). By adopting  such a temperature
reinterpretation, it was possible to keep the entropy of the phantom
field positive definite as required by its probabilistic definition
in the context of statistical mechanics.  In a second approach, a group of authors
\cite{limamaia,LA04} have advocated that the temperature of any dark
energy component is always positive definite obeying the evolution
law 
\beq 
T \sim a^{-3\omega}\,,\label{temp2} 
\eeq 
and, more important, that the existence of phantom fluids is not
thermodynamically consistent because its co-moving entropy is
negative since  $S \sim (1+ \omega)T^{1/\omega}a^{3}$. In this
approach, a possible way to save the phantom regime is to introduce
a negative chemical potential to the fluid \cite{ademirsaulo}, so
that the phantom hypothesis is recovered. A chemical potential has also been recently introduced in the context of dark energy $k$-essence models described in terms of a self interacting complex scalar field \cite{bilic}.

In this note we have the intention to shed some light on this
discussion, by favoring a phantom component with positive temperature,
and, under certain thermodynamic conditions, with positive entropy.

\section{Thermodynamic analysis of dark energy fluids}

For simplicity, let us now consider that the homogeneous and
isotropic FRW universe model is dominated by a separately conserved
dark energy fluid described by the EoS (1).  Following standard
lines (see, for instance, Kolb and Turner \cite{kolb}), the
combination of the first and the second law of thermodynamics
applied to a co-moving volume element of unit coordinate volume and
physical volume $V=a^3$, implies that 
\beq 
TdS=d(\rho V)+ p dV
\equiv d[(\rho+p)V]-V dp\,,\label{kt364} 
\eeq 
where $\rho$ and $p$
are the equilibrium energy density and pressure. The integrability
condition, 
\beqq 
{\partial^2 S \over \partial T \partial V}&=&
{\partial^2 S \over \partial V \partial T}\,.\label{kt365} 
\eeqq
leads to the following relation between the energy density and
pressure ($\rho$ and $p$ depends only on the temperature) 
\beq
dp={\rho+p\over T}dT\,,\label{kt367} 
\eeq 
which also follows
directly from the equilibrium expression for the pressure and energy
density. Substituting (\ref{kt367}) into (\ref{kt364}), we have the
differential entropy definition, 
\beq 
dS = {1\over T} d[(\rho +
p)V]-(\rho + p)V {dT\over T^2}=d\bigg[{(\rho+p)V\over T} + C
\bigg]\,,\label{kt368} 
\eeq 
where $C$ is a constant (from now on fixed to be zero). Therefore, up
to an additive constant, the entropy  per co-moving volume must be
defined by 
\beq 
S \equiv{(\rho + p)\over T}V\,,\label{entropy} 
\eeq
a result that remains valid regardless the number of spatial
dimensions. Actually, for  a multidimensional Universe model, the unique
difference is that instead $V=a^{3}$, one must write $V=a^{n}$ for
$n$-spatial dimensions in the above entropy formula. On the other
hand, if the dark fluid expands adiabatically,  $dS=0$, or
equivalently 
\beq 
d\bigg[{(\rho+p)V\over T}\bigg]=0\,,\label{kt370}
\eeq 
which means that the entropy $S$ per co-moving volume is
conserved. The same definition of entropy follows from the  energy
conservation law, $d(\rho V) + pdV=0$, which can be rewritten as
\beq 
d[(\rho+p)V]=V dp\,.\label{kt369} 
\eeq 
As expected, by
inserting (\ref{kt367}) into (\ref{kt369}) one obtains
(\ref{kt370}). Now, using the equation of state (\ref{eqstate}), we
may write the entropy density on the form 
\beq s\equiv {S\over
V}={\rho+p\over T}={(1+\omega)\rho\over T}\,,\label{kt371} 
\eeq
which defines the entropy density in terms of the temperature for a
dark energy fluid. All the above results are very well known, and
the unique loss of generality comes from the fact that the chemical
potential of the dark energy fluid was assumed to be zero from the
very beginning.

At this point we would like to call attention for a paper published
by Youm \cite{donam} related to the entropy of an Universe with
$n$-spatial dimensions.  He assumed  that Eq. (\ref{kt371})  defines the
temperature in terms of the constant entropy thereby getting the scaling law
(\ref{temp1}).  Later on, this approach was adopted by
Gonz\'alez-D\'{i}az and  C. L. Sig\"{u}enza \cite{gonzalez} giving
origin to the idea of  negative temperature in the phantom regime
($\omega <-1$). This interpretation has been subsequently considered in many different contexts 
(see, for instance, \cite{Pacheco} and Refs. therein).

As it appears, this is a very controversial approach since
(\ref{kt371}) defines the entropy density and not the temperature as
assumed by the quoted authors. In order to discuss  this point  we
recall that (\ref{kt371}) can also be obtained  in a very clear
manner trough the Tisza-Callen axiomatic approach. Actually, by
postulating that the entropy (or the energy) is a homogeneous first
order function of the extensive parameters, $S(\lambda U,\lambda V,
\lambda N)=\lambda S(U, V, N)$, one obtains the so-called Euler
relation \cite{callen}
\beq 
TS = U + pV - \mu N, \label{ER} 
\eeq 
where $\mu$  is the
chemical potential and $U=\rho V$ is the internal energy. Therefore,
if the chemical potential is zero, one obtains (\ref{kt371}).  Note
also that by taking the infinitesimal variation of the Euler
relation (\ref{ER}) and combining with the second law one
obtains the Gibbs-Duhem relation 
\beq 
SdT = Vdp - Nd\mu, 
\label{GB}
\eeq 
which reduces to (\ref{kt367}) when the chemical potential is zero (after
inserting the entropy expression given by the above Euler relation).

This is the basics for any homogeneous substance described by the
standard thermodynamics. More important still for the discussion
here, the temperature as defined by
\beq 
\frac{1}{T} = \left(\frac{\partial{S}}{\partial
U}\right)_{V,N}, 
\label{temp}
\eeq 
is always positive definite for the
equilibrium states. Therefore, if the energy density in the cosmological FRW context is
positive (weak energy condition) one may conclude from
(\ref{entropy}), or directly from (\ref{kt371}), that the entropy for
a phantom fluid ($\omega < -1$) is negative definite, and,
therefore, such a component is thermodynamically forbidden. Note
also that all dark energy fluids with $\omega > -1$ have positive
entropies, a result obtained before the Supernovae observations
\cite{limamaia}. In addition,  once the dependence of the energy
density on the scale factor $\rho(a)$ is established for an
expanding adiabatic Universe, the expression for the entropy itself
determines the temperature evolution law as happens for the cosmic
background radiation ($\omega = 1/3$). Naturally, this approach to
determine the temperature law is not valid if the system evolves
trough a sequence of non-equilibrium states as happens when bulk
viscosity \cite{bulk} or irreversible matter creation \cite{mc}
mechanisms are taken into account. It should also be remarked that
the temperature evolution law can also be obtained even when the hypothesis that the 
energy density and pressure are functions only on the temperature and does not need to be explicitly
used as discussed above. This approach will be discussed in the next section by using only local 
variables in the FRW background. 

\section{Temperature Evolution Law in the FRW geometry}

The equilibrium thermodynamic states of a relativistic simple
fluid obeying the $\omega$-EoS can be completely characterized
by the conservation laws of energy, the number of particles, and entropy.
In terms of specific variables, $\rho$, $n$ (particle number
density) and $s$ (entropy density) the above quoted laws for a FRW type background can be expressed as
\begin{equation}
 \dot{\rho} + 3 (1 + \omega)\rho \frac{\dot a}{a}=0,\,\,
 \dot{n} + 3n\frac{\dot a}{a}=0,\,\, \dot{s} + 3s\frac{\dot a}{a}=0, \label{eqDiff}
\end{equation}
with general solutions of the form:
\begin{equation}
\rho=\rho_0 \left(\frac{a_0}{a} \right)^{3(1 + \omega)}, \,\, n=n_0
\left(\frac{a_0}{a}\right)^{3}, \,\, s=s_0
\left(\frac{a_0}{a}\right)^{3},\label{eqSol}
\end{equation}
where $\rho_0$, $n_0$, $s_0$ and $a_0$ are present day (positive) values of the
corresponding quantities. On the other hand, the
quantities $p$, $\rho$, $n$ and $s$ are related to the temperature
$T$ by the Gibbs law
\begin{equation} \label{eq:GIBBS}
nTd\big({s\over n}\big)= d\rho - {\rho + p \over n}dn,
\end{equation}
and from the Gibbs-Duhem relation (\ref{GB}) there are only two
independent thermodynamic variables, say $n$ and $T$. Therefore,
by assuming that $\rho=\rho(T,n)$ and $p=p(T,n)$, one may show that the
following thermodynamic identity must be satisfied 
\begin{equation}
T \biggl({\partial p \over \partial T}\biggr)_{n}=\rho + p - n
\biggl({\partial \rho \over \partial n}\biggr)_{T},
\end{equation}
an expression that remains locally valid even for out of equilibrium states \cite{weinb}. 
Now, inserting the above expression into the energy conservation law as
given by (\ref{eqDiff}) one may show that the temperature satisfies
\begin{equation} \label{eq:EVOLT}
{\dot T \over T} = \biggl({\partial p \over \partial
\rho}\biggr)_{n} {\dot n \over n} = -3\omega \frac{\dot a}{a},
\end{equation}
and assuming $\omega \neq 0$ a straightforward integration yields
\begin{equation} \label{eq:TV}
n = n_0 \left(\frac{T}{T_0} \right)^{1 \over  \omega}  \quad
\Leftrightarrow \quad T=T_0 \left(\frac{a}{a_0} \right)^{-3\omega}.
\end{equation}
In the standard fluid description,  the temperature appearing in the
above expressions is positive regardless of the value of $\omega$.
Note also that the temperature evolution law is completely independent of the
entropy density. The above expressions also imply that for a given
co-moving volume $V$ of the fluid, the product $T^{1 \over  \omega}
V$  remains constant and must characterize the equilibrium states
(adiabatic expansion). At this point, the above temperature law, $T \sim a^{-3\omega}$, should be 
compared with the one proposed in Refs. \cite{donam,gonzalez}, namely,
$T \sim (1+\omega)a^{-3\omega}$. It shows that the  prefactor ($1+\omega$) 
in the temperature law is completely artificial, and, therefore, it has no physical meaning. 
Moreover, the entropy expression as  given by the 
Euler relation ($\ref{entropy}$) with $\mu=0$, is just telling us that the 
phantom fluid is thermodynamically forbidden because the entropy of a dark energy fluid becomes negative for $\omega < -1$. 

In an attempt to turn acceptable a phantom fluid with negative temperature, the authors of Ref. 
\cite{gonzalez} comment on some quantum
mechanical systems with negative temperatures. Actually, the possibility of negative values
of temperature has been discussed by several authors
\cite{ramsey,landsberg59,bloembergen}. From Eq. (\ref{temp}) 
one may conclude that the temperature may be
negative if the entropy diminishes while the internal energy grows. 
This may happens, for instance, in some condensed matter system when the energy spectrum is 
limited from above thereby presenting population inversion phenomenon as required for the
operation of semiconductor lasers \cite{landsberg}. 
Such an effect for paramagnetic
systems of nuclear moments in a crystal were studied in detail by Purcell and
Pound \cite{purcell}. However, as remarked by  Izquerdo and Pav\'on \cite{pavon06},
all models of phantom energy models proposed so
far in literature assume some type of scalar field with no upper
bound on their energy spectrum. Moreover, while population inversion
is a rather transient phenomenon, the phantom regime is supposed to last for many eons. 
In a point of fact, bodies of negative temperature would
be completely unstable and in principle they cannot exist 
naturally in the Universe, except in some singular states of a system \cite{landau}. Such
states are out of equilibrium (different from the analysis assumed in 
Refs. \cite{donam,gonzalez}). They can be produced only in
certain very unique systems, specifically in isolated spin systems,
and they spontaneously decay away \cite{callen}.

The considerations presented  in the  two previous sections may induce someone to 
think that phantom fluids cannot exist in nature or that the statistical mechanics 
and thermodynamics need to be somewhat generalized, as for instance, by adopting the 
non-extensive framework proposed by Tsallis \cite{tsallis}. However, it should be recalled that all the results  above discussed are valid only if the chemical 
potential of the phantom fluid is identically zero.

\section{Saving the Phantom Hypothesis}

As we have argued,  the concept of negative temperature
is not a reasonable physical or mathematical solution to save the phantom hypothesis. Therefore, the important question now is 
how a phantom fluid may exist with temperature and entropy
positives. In principle, it should be nice if such a problem might be solved in the framework of the standard 
thermodynamics and statistical mechanics. 
As far as we know, the unique possibility available to us is to
introduce a new thermodynamic degree of freedom, namely, the
chemical potential, a quantity appearing naturally  in the Euler and Gibbs-Duhem relations. 

For  one component fluid the Gibbs free energy $G$ is defined by:
\beq
G(T,p,N)\equiv U+pV-TS\,,
\eeq
with differential
\beq
dG= -SdT + Vdp + \mu dN\,,
\eeq
yielding for the chemical potential 
\beq
\mu = \bigg({\partial G\over \partial N}\bigg)_{p,T}\,.
\eeq
Now, by using relations (\ref{eqstate}), (\ref{eqSol}) and (\ref{eq:TV}) it is straightforward to show that
\beq
G=[(1+\omega){\rho_0\over n_0}{T\over T_0} - {s_0\over n_0} T]N,
\eeq
and
\beq\label{muf}
\mu=\bigg[{(1+\omega)\rho_0-s_0T_0\over n_0}\bigg]{T\over T_0}\equiv \mu_0{T\over T_0}\,,
\eeq
where we have defined $\mu_0\equiv [(1+\omega)\rho_0-s_0T_0]/n_0$ as the present day value of the chemical potential. We see that, for the phantom regime ($\omega <-1$), we have $\mu_0 <0$ as well as $(\partial \mu/\partial T)<0$. Moreover, for $T\to 0$, we have that $\mu$ increases toward zero. 

Therefore, if $\mu$ is different from
zero, one may show that the entropy (\ref{entropy}) must be
replaced by (see also Eq. (\ref{ER}))
\beq S(T,V) = \left[\frac{(1+\omega)\rho_0-\mu_0
n_0}{T_0}\right] \bigg({T\over T_0}\bigg)^{1/\omega}V\,.\label{mu}
\eeq 
The basic conclusion here is that in order to
keep the entropy positive definite, the following constraint must be
satisfied \cite{ademirsaulo}: 
\beq \omega \geq \omega_{min}=-1 +
{\mu_0 n_0 \over \rho_0}\,,\label{accord} 
\eeq 
which introduces a
minimal value to the $\omega$-parameter below which the entropy
becomes negative. Therefore, only in this case a dark energy 
component in the phantom regime ($\omega_{min} < -1$) is 
endowed with a positive  entropy as required from 
Boltzmann's microscopic definition ($S=k_B lnW$). 

It is also worth noticing that such a
condition can also be obtained by a completely different approach,
namely, by studying  the accretion of a phantom fluid with
non-zero chemical potential by a black hole with basis on the
generalized second law of thermodynamics \cite{LPJG}. 
Still more important, since the chemical potential of an ideal relativistic bosonic gas satisfies $\mu \leq mc^{2}$,  we see that Eq. (\ref{muf}) plus the entropy constraint  ($S>0$) suggest a bosonic massless nature to the phantom particles.

\section{Concluding remarks}

Contrary to the claims of some authors \cite{gonzalez}, we have shown 
here that the temperature of a dark energy fluid must be always 
positive definite in the range $\omega > -1$, and that the phantom  regime ($\omega <-1$) 
is thermodynamically forbidden, because its entropy is negative. Based on a 
straightforward thermodynamic analysis of the dark energy regime in the 
FRW metric we have demonstrated that the true thermodynamic temperature 
evolution law is given by the form $T \sim a^{-3\omega}$ as previously 
derived \cite{limamaia,LA04}.  Finally, we have also advocated that in order to give some physical 
meaning to the phantom regime one needs to include  a negative chemical potential. Only in this case, the entropy of the phenomenological phantom fluid hypothesis 
is not in contradiction with the probabilistic definition of the thermodynamic entropy.

\begin{acknowledgments}

The authors would like to thank V. C. Busti, J. V. Cunha, J. F.
Jesus, A. C. Guimar\~aes, R. Holanda and R. C. Santos for helpful
discussions. JASL is partially supported by CNPq and FAPESP (Brazilian
Research Agencies) under Grants 304792/2003-9 and 
04/13668-0, respectively. SHP is supported by CNPq No. 150920/2007-5. 
\end{acknowledgments}

\end{document}